\documentclass[fleqn,usenatbib]{mnras}

\usepackage{newtxtext,newtxmath}
\usepackage[T1]{fontenc}
\usepackage{ae,aecompl}
\usepackage{times}
\usepackage[export]{adjustbox}

\usepackage{rotating,graphicx}	
\usepackage{amsmath}	
\usepackage{amssymb}	
\usepackage{ulem}
\usepackage[ampersand]{easylist}

\usepackage{latexsym}
\usepackage{enumerate}
\usepackage{amstext}
\usepackage{morefloats}

\usepackage{natbib}
\usepackage{rotating}
\usepackage{changepage}
\usepackage{scrextend}
\usepackage[ampersand]{easylist}
\usepackage{xcolor}
\usepackage{ulem}

\newcommand{\rt}[1]{{\textcolor{black}{{#1}}}}
\newcommand{\btt}[1]{{\textcolor{black}{{#1}}}}

\newcommand{\figdir}{./}
\usepackage{mathtools}

\title[Gas in Globular Clusters: Winds and Pulsar Heating]{Modeling Gas Evacuation Mechanisms in Present-Day Globular Clusters: Stellar Winds from Evolved Stars and Pulsar Heating}

\author[J.~P.~Naiman et al.]{J~P.~Naiman$^{1}$\thanks{E-mail: jill.naiman@cfa.harvard.edu}, M. Soares-Furtado$^{2}$, E. Ramirez-Ruiz$^{3}$
\vspace*{0.2cm} \\
$^{1}$Harvard-Smithsonian Center for Astrophysics, 60 Garden Street, Cambridge, MA 02138, USA\\
$^{2}$Department of Astrophysical Sciences, Princeton University, Princeton, NJ 08544, USA\\
  $^{3}$Department of Astronomy and Astrophysics, University of California, Santa Cruz, CA 95064, USA
}

\date{11 July 2017}

\pubyear{2018}

\begin{document}
\label{firstpage}
\pagerange{\pageref{firstpage}--\pageref{lastpage}}
\maketitle

\begin{abstract} 
We employ hydrodynamical simulations to investigate the  underlying mechanism responsible for the low levels of gas and dust in globular clusters.
Our models examine the competing effects of energy and mass supply from the various components of the evolved stellar population for globular clusters 47 Tucanae, M15, NGC~6440, and NGC~6752.
Ignoring  all other gas evacuation processes, we find that the energy output from the stars that have recently turned off the main sequence are capable of effectively  clearing the evolved stellar ejecta and producing intracluster gas densities consistent with current observational constraints.
This result distinguishes a viable gas and dust evacuation mechanism that is ubiquitous among globular clusters.
In addition, we extend our analysis to probe the efficiency of pulsar wind feedback in globular clusters.
We find that if the energy supplied by the pulsar winds is effectively thermalized within the intracluster medium, the material would become unbound.
The detection of intracluster ionized gas in 47~Tucanae allows us to place particularly strict limits on pulsar wind thermalization efficiency, which must be extremely low in the cluster's core in order to be in accordance with the observed   density constraints.
\end{abstract}




\section{Introduction}

For over half a century, globular cluster observations have revealed a paucity of intracluster dust and gas.
Given the abundance of stellar ejecta supplied by a population of evolved stars and the extensive timescales between Galactic disk crossing events, these observations are at odds with theoretical expectations. Orbiting in the Galactic halo, globular clusters traverse through the plane of the galaxy on timescales of ${\sim}10^{8}$ years, expelling the intracluster medium with each passage \citep{Odenkirchen1997}.
Between Galactic disk crossing events, the evolving stellar members continuously fill the cluster with stellar ejecta. These stars are about \rt{$0.8 - 0.9 \, {\rm M_\odot}$} at the main sequence turn-off.  During the evolution to the white dwarf stage, 10~$\rm M_{\odot}$ -- 100~$\rm M_{\odot}$ of material is predicted to have been accumulated \citep{Tayler1975}. The hunt for this elusive intracluster medium has been extensive, yet the majority of observations have been fruitless.  Searches for dust as well as atomic, molecular, and ionized gas have been carried out, resulting in upper limits and detections  that are generally much lower than the values expected if the mass was effectively retained.

Submillimeter and infrared (IR) searches for dust in globular clusters \citep{Lynch1990, Knapp1995, Origlia1996, Hopwood1999} have predominately resulted in low upper limits for the dust mass content in the systems.
For example, measurements by \cite{Barmby2009} found an upper limit of $\leq 4\times 10^{-4} \, {\rm M_{\odot}}$. This can be compared to theoretically predicted dust masses for globular clusters, which range from $10^{-3.6} {\rm M_\odot}$ -- $10^{-0.8}{\rm M_\odot}$.
Tentative evidence for excess IR emission from cool dust in the metal-rich globular cluster NGC~6356 was found nearly two decades ago \citep{Hopwood1998}, although the lack of a 90~$\mathrm{\mu}$m excess casts doubt on this detection \citep{Barmby2009}. Only the Galactic globular cluster M15 (NGC~7078) depicts clear evidence for an IR  excess \citep{Evans2003, Boyer2006}, revealing a cluster dust mass  of $9\pm2\times 10^{-4} \, {\rm M_\odot}$, which is at least one order of magnitude below the predicted value.  The lack of intracluster dust might suggest that evolved stars  produce less dust than predicted, although evidence to the contrary has been found in  M15 and NGC~5139, where some of the dustiest, most mass-expelling stars have been found \citep{Boyer2006,Boyer2008}.

\citet{Troland1978} led the first search for 2.6~mm CO emission in globular clusters, which resulted in a non-detection, but lacked sufficient sensitivity  to rule out the presence of molecular gas. Since then, upper limits have been placed on the molecular gas content in globular clusters, constraining the mass to about 0.1~$\rm M_{\odot}$ \citep{Smith1995, Leon1996}. The most promising search lead to a tentative  detection of two CO lines in the direction of globular cluster 47 Tucanae (NGC~104), which was interpreted to be the result of the bow shock interaction generated as the cluster traverses through the Galactic halo \citep{Origlia1997}. Other searches, including attempts to measure OH and $\rm H_{2}O$ maser emission, have been unsuccessful \citep{Knapp1973a, Kerr1976, Frail1994, Cohen1979, Dickey1980, vanLoon2006}.

Neutral hydrogen (HI) at 21~cm was detected in NGC~2808, measuring 200~$\rm M_{\odot}$ of gas \citep{Faulkner1991}. This is not beyond dispute, however, since there is known to be a foreground 21~cm extended region around the cluster. Most other attempts to detect HI in globular clusters have been unsuccessful or resulted in upper limits on the order of a few solar masses \citep{Heiles1966, Robinson1967, Kerr1972, Knapp1973b, Bowers1979, Birkinshaw1983, Lynch1989, Smith1990, vanLoon2006, vanLoon2009}. A tentative HI detection of 0.3~$\rm M_{\odot}$ in M15 was presented by \citet{vanLoon2006} using 21~cm Arecibo observations.

The  most reliable intracluster medium constraints are derived from radio dispersion measurements of  known millisecond pulsars in  47~Tucanae \citep{Camilo2000}, which resulted in the first detection of ionized  gas with a density of $n_{\rm e}=0.067 \pm 0.015\; {\rm cm}^{-3}$  \citep{Freire2001b}. This ionized gas measurement, as well as the upper limits placed on ionized gas in other clusters \citep{Smith1976, Faulkner1977, Knapp1996}, corresponds to a gas deficiency of two to three orders of magnitude when contrasted with the amount of stellar ejecta predicted to accumulate within the cluster.

 This dearth of gas appears to be ubiquitous among globular clusters and suggests a common mechanism acting to continuously expel gas from the environment. Potential gas evacuation processes may be external or internal to the globular cluster.
Ram pressure stripping is an external process, produced as the globular cluster traverses through the surrounding hot Galactic halo.
This mechanism has been investigated both analytically and numerically in the past. \citet{Frank1976} found that the interstellar medium of the Galactic halo was too low in density by one order of magnitude to account for stripping of the intracluster medium.   \citet{Priestley2011}, aided by the use of  three-dimensional hydrodynamical simulations, revisited this problem and concluded  that halo sweeping was only an effective gas evacuation mechanism for globular clusters with $M \leq 10^5 \, {\rm M_{\odot}}$. This is further compounded by the fact that the majority of  globular clusters reside in low density regions of the halo.

Internal evacuation mechanisms are more varied in scope with some being impulsive and others being continuous in nature.
\citet{Vandenberg1977} suggested the possibility that UV heating from the horizontal branch (HB) stellar population might provide sufficient energy input to explain low gas densities in clusters, however not all clusters contain hot HB stars. \citet{Umbreit2008} argues that the energy injected by stellar collisions could be significant, in particular in clusters with high encounter rates.  \citet{Coleman1977} investigated the possibility that flaring M-dwarf stars might supply the energy injection required to evacuate the cluster, however the number and distribution of M-dwarf stars in clusters is highly uncertain. The energy injection from hydrogen rich novae explosions is another possible gas evacuation mechanism, which was investigated early on by \citet{Scott1978} and more recently by  \citet{Moore2011}. However, it is highly uncertain whether these explosions occur with enough frequency  \citep{Bode2008}  and  if the beamed structure of the emanating outflows will lead to significantly lower gas removal  efficiencies \citep{Obrien2006}.  Heating by white dwarfs may be an effective gas removal process, but generally only for low mass and low stellar density clusters \citep{mcdonald2015}.
Finally, the presence of millisecond pulsars in many of these systems not only   enables the placement of stringent intracluster density constraints, but also provides globular clusters with yet another mechanism  of energy injection \citep{Spergel1991}.

Among the gas evacuation mechanisms described above, the internal energy injection processes described in the preceding paragraph are expected to vary significantly between clusters such that, individually, they would be unable to explain the universality of low gas densities seen across all globular clusters.
Mass and energy injection from stellar winds, on the other hand,  are  a  common feedback ingredient  in all clusters.
In the past,  it has been  argued that the evolved stellar ejecta does not  posses sufficient energy to escape the cluster potential \citep{Vandenberg1977}.
This, however,  has been called into question by observations of giant stars with wind velocities exceeding the typical cluster escape velocity \citep{Smith2004,dupree2009}.
What is more, energy injection from the usually neglected, although plentiful, stars recently turning off the main sequence could play a decisive role in mediating mass retention in these systems as they produce an energy injection comparable to that of main sequence winds \citep{Smith1999}.

Motivated by this line of reasoning, we present a systematic investigation of the impact of outflows emanating from stars evolving off the main sequence to the RGB and AGB on the intracluster gas evolution.
To aid in our interpretation of the data, we compare observational constraints with the results of hydrodynamical simulations.  These simulations include radiative cooling as well as mass and energy injection  from  the cluster members, which we  derive using  stellar evolution models.
It is shown that the observational constraints of intracluster gas can be successfully explained in models where the diffuse hot stellar winds emanating from numerous recent turn-off stars efficiently thermalizes with the slow and massive winds of the few highly evolved RGB/AGB stars in a cluster\rt{ -- a process by which the entire gas content of the cluster is effectively heated}.
In globular clusters with stringent gas density constraints, such as 47~Tucanae, M15, NGC~6440, and NGC~6752, we argue that energy output from the recent turn-off stellar population alone is  capable of efficiently  clearing out the  evolved stellar ejecta.
Since the majority of clusters with stringent gas  content constraints host millisecond  pulsars, we extend our calculations to include the energy injection  supplied by their  winds. In particular, the detection of ionized gas in 47~Tucanae allows us to place strict limits on the pulsar wind thermalization efficiency within these systems.

\begin{figure*}
\centering\includegraphics[width=0.99\textwidth]{\figdir 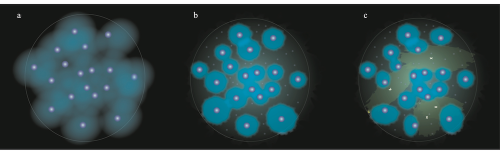}
\caption{ Diagram illustrating the mass and energy contributions arising from evolved red giant winds, turn-off stars, and pulsar winds.
If the evolved RGB stars dominate the mass and energy injection, a lower limit on the gas density can be calculated by assuming that their emanating winds extend only to their closest neighbors (panel {\it a}).
While the evolved RGB stellar members are expected to dominate the mass injection, despite comprising a small subset of the stellar cluster population,  the energy injection is likely  to be dominated  by the more abundant recent turn-off stars (panel {\it b}). Here the extension of the stellar winds is confined to a smaller volume around the RGB members.
In some clusters, millisecond pulsar winds (positioned at the \textit{x} points in panel {\it c}) provide an additional energy injection source.
A significant amount of  energy  injection could  prevent the winds from the most evolved stellar members from effectively expanding between the closest neighbors, resulting in lower gas content. This is illustrated in panels {\it b} and {\it c} when the energy injection is dominated by recent turn-off stars and pulsar winds, respectively.}
\label{fig:diagram}
\end{figure*}

\section{Modeling Gas in Globular Clusters}

\btt{Many authors have pointed out the discrepancy between observed levels of intracluster gas and that predicted by the full retention of red giant ejecta \cite[e.g.][]{Tayler1975}.  Typical estimates are calculated assuming winds emanating from each star fill the spaces between stars and halt their expansion, or the gas from each star mixes and expands to fill the entire cluster volume.  Following such work as  \cite{Pfahl2001}, we} begin with a simple estimate of the intracluster gas density considering a cluster comprised of
$N=10^6$ stars with masses of $0.9 \, {\rm M_\odot}$ at the main sequence turn-off. The ratio of evolved stars to the total number of stars within the cluster is roughly $N_{\rm to}/N \approx 0.01$.
\btt{Assuming the volume of the cluster can be approximated by the sum of the individual separation between stars - $N r_\perp^3 \approx r_{\rm h}^3$, t}his results in an estimated average separation between the evolved stars that is equal to
\begin{equation}
r_\perp=6.4 \times 10^{17} \left(\frac{N_{\rm to}}{10^2}\right)^{-1/3} \left(\frac{r_{\rm h}}{1{\rm pc}}\right)\;{\rm cm},
\end{equation}
where $r_{\rm h}$ represents the cluster  half-light radius.
We estimate a lower and upper limit for the expected cluster density, ignoring the effects of intracluster gas evacuation mechanisms.
\btt{Following \cite{Pfahl2001}, i}f the winds from each evolved stellar member extend only to its nearest neighbors an estimate of the cluster gas density is found to be
\begin{equation}
n_\perp \approx \frac{3 \dot{M}_{\rm w, to} \Delta t_\perp}{4 \pi r_\perp^3 v_{\rm w, to}}
\end{equation}
\btt{where $\dot{M}_{\rm w,to}$  and  $v_{\rm w,to}$ correspond to the mass-loss rate and wind velocity of the evolved stellar members (Section~\ref{sec:se} contains a more detailed description of stellar evolution models). Assuming the time for the winds to spread to the distance of the evolved star separation, $r_\perp$, is simply $\Delta t_\perp = r_\perp/v_{\rm w,to}$ this equation can be expressed simply as}
\begin{equation}
n_\perp=0.1 \left(\frac{N_{\rm to}}{10^2}\right)^{2/3} \left(\frac{r_{\rm h}}{1{\rm pc}}\right)^{-2} \left(\frac{v_{\rm w,to}}{70\;{\rm km\;s^{-1}}}\right)^{-1} \left(\frac{\dot{M}_{\rm w,to}}{10^{-7}\, {\rm M_\odot}\;{\rm yr^{-1}}}\right)\,{\rm cm}^{-3},
\end{equation}

Certainly, we expect continued mass loss from the population of evolving stars, so this estimate is intended to provide a lower limit expectation of the gas density for a system devoid of gas evacuation mechanisms.
This lower limit may then be contrasted with a second estimate, where evolved stellar winds extend to fill the volume of the entire cluster.
\btt{In this case the density can be expressed as}
\begin{equation}
n_{\rm f} = \frac{3 N_{\rm to} \dot{M}_{\rm w, to} \Delta t_{\rm c}}{4 \pi r_{\rm c}^3}
\end{equation}
\btt{with $\Delta t = r_{\rm c}/v_{\rm w,to}$} the density for this second scenario is found to be
\begin{equation}
n_{\rm f}=1.5  \left(\frac{N_{\rm to}}{10^2}\right) \left(\frac{r_{\rm h}}{1{\rm pc}}\right)^{-2} \left(\frac{v_{\rm w,to}}{70\;{\rm km\;s^{-1}}}\right)^{-1} \left(\frac{\dot{M}_{\rm w,to}}{10^{-7}\, {\rm M_\odot}\;{\rm yr^{-1}}}\right)\,{\rm cm}^{-3}.
\end{equation}
In fact, we suspect that the gas density may exceed $n_{\rm f}$  if $v_{\rm w, to} \lesssim \sigma_{\rm v}$, where $\sigma_{\rm v}$ is the velocity dispersion of the cluster, when the gravitational effects of the cluster are taken into consideration \citep{Pflamm2009}. \btt{Both these upper and lower estimates on the density expected from evolved stellar members in present-day globular clusters result in significantly higher gas densities than those observed  \citep[e.g.][]{Knapp1995}.}

On the other hand, if significant energy injection takes place within the cluster,  the gas density is expected to be lower than the lower gas density limit  $n_\perp$.  The impact of energy injection on the intracluster gas content is presented in Figure~\ref{fig:diagram} for three illustrative cases: the emanation of cool winds from a small population of highly evolved stars at the end of the AGB and RGB branches (panel {\it a}), a mix of this small population of cool wind generators with the hot winds from an abundant recent turn-off population (panel {\it b}) and from the combined effort of the cool and hot stellar ejecta and millisecond pulsar winds (panel {\it c}).

\btt{While this diagram shows qualitatively the effects of extra energy injection in globular clusters on the retained gas content, more precise determinations of intracluster densities require numerical simulation.}
Much of our effort in this paper will  be dedicated to determining the state of the intracluster gas  in various globular clusters, and describing how the expected energy injection from stars in various stages of post-main sequence evolution and millisecond pulsar populations may affect mass retention in these systems.

\section{Numerical Methods and Initial Setup}\label{sec:num}

\subsection{Hydrodynamics}
Our investigation of evolved stellar ejecta retention within globular clusters is mediated by hydrodynamical simulations incorporating energy and mass injection. For the purpose of simplicity, spherical symmetry is assumed \citep{Quataert2004, Hueyotl2010}. {\tt FLASH},  a parallel, adaptive-mesh hydrodynamical code \citep{Fryxell2000}, is employed to solve the hydrodynamical equations in one-dimension. Within the cluster, emanating winds from the dense stellar population and millisecond pulsars, when present, are assumed to shock and thermalize and, as such, the mass and energy contributions are implemented as source terms in the hydrodynamical equations.
In spherical symmetry, the hydrodynamical equations may be written following the simplified forms presented in \cite{Naiman2012} as:
\begin{equation}
\rho(r,t_n) = \rho(r,t_{n-1}) + q_{\rm m,\star}(r,t_{n}) dt(t_n,t_{n-1})
\label{eq:masscons}
\end{equation}
\begin{equation}
u(r,t_{n}) = u(r,t_{n-1}) \rho(r,t_{n-1})/\rho(r,t_{n}) + a_g(r,t_{n+1})dt(t_n,t_{n-1})
\label{eq:momentum}
\end{equation}
\begin{equation}
\begin{multlined}
\rho(r,t_n)\varepsilon(r,t_n) = \frac{1}{2}\rho(r,t_{n-1})u(r,t_{n-1})^2 + \rho(r,t_{n-1})\varepsilon(r,t_{n-1}) \\
- \frac{1}{2}\rho(r,t_{n})u(r,t_{n})+ q_{\varepsilon,\star}(r,t_{n-1})dt(t_n,t_{n-1})\\
 + q_{\varepsilon,\Omega}(r,t_{n-1})dt(t_n,t_{n-1}) -Q(r,t_{n-1})
\label{eq:energy}
\end{multlined}
\end{equation}
where $\rho(r,t)$, $u(r,t)$ and $\varepsilon(r,t)$ correspond to the gas density, radial velocity and internal energy density, respectively  \citep{Holzer1970,Hueyotl2010}.
$Q(r,t) = n_{\rm i}(r,t) n_{\rm e}(r,t) \Lambda(T,Z)$ is the cooling rate for a gas consisting of ion and electron number densities, $n_{\rm i}(r,t)$ and $n_{\rm e}(r,t)$, and the cooling function for gas of temperature $T$ with metallicity $Z$ is represented by $\Lambda(T,Z)$.
Cooling functions are taken from \cite{Gnat2007} for $T > 10^4 \, {\rm K}$ and
from \cite{Dalgarno1972} for $10 \, {\rm K} \le T \le 10^4 \, {\rm K}$.  Following \cite{Naiman2012}, we do not include any additional energy or momentum terms due to the motion of stars within the cluster or the non-radially-uniform distribution, and thus mass and energy deposition, of the stellar members.

In equations (\ref{eq:masscons})-(\ref{eq:energy}), the terms $q_{\rm m,\star}(r,t)$ and $q_{\varepsilon,\star}(r,t)$ respectively represent the rates of mass and energy injection produced by the evolved stellar ejecta at a time $t$ in a cluster's history.  \btt{Here, $dt(t_n,t_{n-1})$ is the timestep between simulation times $t_{n-1}$ and $t_n$.}
Stellar winds dominate cluster mass injection, and, as a result, the hydrodynamical influence of the millisecond pulsars is  restricted to the energy injection term: $q_{\varepsilon,\Omega}(r,t)$.
Given $N$ stars, each with an average mass-loss rate (at a particular evolutionary  time $t$) of  ${\langle \dot{M}(t)\rangle}$ and a wind energy injection rate ${\frac{1}{2} \langle \dot{M}(t)\rangle \langle v_{\rm w}(t)^2\rangle}$, we find a total mass-loss of ${\dot{M}(t)_{\rm w,total} =  N \langle \dot{M}(t)\rangle = \int 4 \pi r^2 q_{\rm m, \star}(r,t) dr }$ and a total wind energy injection of
$\dot{E}(t)_{\rm w, \star, total} = \frac{1}{2} N \langle \dot{M}(t) \rangle \langle v_{\rm w}(t)^2\rangle = \int 4 \pi r^2 q_{\varepsilon,\star}(r,t) dr$,
where ${q_{\varepsilon,\star}(r,t) = \frac{1}{2} q_{\rm m, \star}(r,t) \langle v_{\rm w}(t)^2 \rangle}$.
In order to preserve simplicity, we have ignored the effects of mass segregation. In addition, we assume
${q_{\rm m,\star}(r,t) \propto n_{\star}(r)}$, such that ${q_{m, \star}(r,t) = A(t) r^{-2} \frac{d}{dr} \left(r^2 \frac{d\Phi_g}{dr}\right)}$, where ${A(t) = \langle \dot{M}(t) \rangle / (4 \pi G \langle M_\star \rangle)}$ and $\langle M_\star \rangle$ corresponds to the average mass of a star.

The stellar cluster gravitational potentials are simulated with a Plummer model, which takes the form
\begin{equation}
\Phi_g = - \frac{G M_{\rm c}}{\left[r^2 + r_{\rm c}^2(\sigma_{\rm v}) \right]^{1/2}}
\end{equation}
for a total cluster of mass $M_{\rm c}$ with velocity dispersion ${\sigma_{\rm v} = \left(3^{3/4}/ \sqrt{2}\right)^{-1} \sqrt{G M_{\rm c}/r_{\rm c}}}$ \citep{Bruns2009, Pflamm2009}.
It should be noted that while the shape of the potential can impact the radial distribution of gas within the core \citep{Naiman2011}, the total amount of gas accumulated within the cluster is relatively unaffected by the shape of the potential.
In addition, to account for the gas dynamics under the influence of $\Phi_g$,  the self gravity of the gas is computed using {\tt FLASH}'s multipole module.  The resolution is fixed to 6400 radial cells for each model. The core radius, $r_{\rm c}$,  sets the resolution within the computational domain, ensuring that we adequately resolve the
core and setting a cluster potential of effectively zero at the outer boundary.  As the cell centers range from approximately $r_{\rm min} \sim 0.08 r_{\rm c}$ to $r_{\rm max} \sim 250 r_{\rm c}$, we employ an exponential fall off in the potential at a scale length of 10~pc at the tidal radius of a cluster, as approximated by $r_t = R_{\rm gc} \left( M_{\rm c} / 2 M_{g} \right) ^{1/3}$ \rt{for a Galactic potential with a constant circular rotation curve,} where the mass of the Milky Way is taken as $M_g = 6.8 \times 10^{11} \, {\rm M_\odot}$ \citep{eadie2016} and the galactocentric distance to the cluster is given by $R_{\rm gc}$ (Table \ref{table:clusterparams}).  All models are run until they reach steady state, until the simulation time reaches the age of the cluster, or star formation is triggered following the prescription discussed in \cite{Naiman2012}.
Moderate changes in domain size or resolution do not affect our results.  However, it is worth noting that domains which are small enough to not allow gas injected by stellar winds to escape the cluster potential can lead to artificially large central density enhancements as gas that would otherwise escape is funneled into the center of the cluster.

\subsection{Stellar Evolution}\label{sec:se}

The gas evolution resulting from simulated stellar winds are highly dependent upon two key parameters: the time dependent average stellar mass-loss rate and stellar wind velocity. The average stellar mass-loss rate and stellar wind velocity then directly determine the mass and energy injection rates, $q_{\rm m,\star}$ and $q_{\varepsilon,\star}$. Because we are employing spherically symmetric simulations, these rates must encompass the average mass-loss properties of the stellar population as a whole.
Following \cite{Naiman2012}, we derive the average mass-loss rate and wind velocity for a population of stars by integrating wind properties over their initial IMF and formation history \citep{Kroupa2013}.

The average mass $\langle \Delta M(t_i)\rangle$  and kinetic energy  $\langle \Delta E_K(t_i) \rangle$  injection  within the cluster at  a time $t_i$  by a population of stars of $M_\star \epsilon \;[M_{\rm L}, M_{\rm H}]$ born as a single stellar population at time $t_0$ may then be formalized \citep{Naiman2012}:
 \begin{equation}
\langle \Delta M(t_i) \rangle = \int_{t_0}^{t_i} \int_{M_{\rm  L}}^{M_{\rm  H}}
\zeta(M_\star,t) \dot{M}(M_\star,t)  dM_\star dt
\label{deltam}
\end{equation}
and
 \begin{equation}
\langle \Delta E_K(t_i) \rangle = \frac{1}{2} \int_{t_0}^{t_i}  \int_{M_{\rm  L}}^{M_{\rm  H}}
\zeta(M_\star,t) \dot{M}(M_\star,t) v_w^2(M_\star,t) dM_\star dt,
\label{deltae}
\end{equation}
respectively.  Here $\zeta(M_\star,t) = \zeta(M_\star)$ is the non-evolving IMF, assumed to be accurately described by the \citet{kroupa2001} IMF, and $\dot{M}(M_\star,t)$ and $v_{\rm w}^2(M_\star,t)$ are the mass-loss rates and wind velocities for individual stars of mass $M_\star$, respectively.

Stars with lifetimes  $t_{\rm age}(M_\star) < t_i$ abandon the stellar population and are not included in the averaging.  For any given time we denote stars turning off the main sequence at $t_i = t_{\rm to}$, with masses $M_{\rm H} = M_{\rm to}$, permitting us to split these equations into their corresponding turn-off and main sequence components. These are given by:
\begin{equation}
\begin{multlined}
\langle \Delta M(t_i) \rangle =  \langle \Delta M_{\rm to} \rangle + f_{\rm ms} \langle \Delta M_{\rm ms} \rangle \\
= \left( \frac{N_{\rm to}}{N_{\rm tot}} \right)\int_{t_0}^{t_i} \dot{M}(M_{\rm to},t) dt \\
+   f_{\rm ms} \int_{t_0}^{t_i}  \int_{M_{\rm  L}}^{M_{\rm  H} < M_{\rm to} (t_i)}
\zeta(M_\star) \dot{M}(M_\star,t)  dM_\star dt
\label{deltam2}
\end{multlined}
\end{equation}
and
\begin{equation}
\begin{multlined}
\langle \Delta E_K(t_i) \rangle = \langle \Delta E_{K, {\rm to}} \rangle  + f_{\rm ms} \langle \Delta E_{K, {\rm ms}} \rangle\\
= \frac{1}{2}  \left( \frac{N_{\rm to}}{N_{\rm tot}} \right) \int_{t_0}^{t_i} \dot{M}(M_{\rm to},t) v_{\rm w}^2(M_{\rm to},t)dt \\
 + \frac{f_{\rm ms}}{2} \int_{t_0}^{t_i}  \int_{M_{\rm  L}}^{M_{\rm  H} < M_{\rm to} (t_i)}
\zeta(M_\star) \dot{M}(M_\star,t) v_{\rm w}^2(M_\star,t) dM_\star dt,
\label{deltae}
\end{multlined}
\end{equation}
where the $to$ subscripts denote the calculated quantities are from the turn-off star \rt{{\textbf{alone}}}, the fraction of the main sequence stellar winds that is effectively thermalized and mixed within the cluster environment is denoted by $f_{\rm ms}$, and the ratio of the number of turn-off stars to the total number of stars in the stellar population is given by $\left(N_{\rm to}/N_{\rm tot}\right)$.  In \cite{Naiman2012} mass retention and expulsion in star clusters for the two cases of $f_{\rm ms} = 0$ and $f_{\rm ms} = 1$ were explored.  In what follows, we take $f_{\rm ms} = 0$ and instead focus on mass-loss from stars with $M_\star \sim M_{\rm to}$ at different phases in their evolution off the main sequence towards the RGB and AGB phases to test whether evolved stellar ejecta alone can explain the dearth of gas in present day globular clusters.  This allows for quantification of the effects of mixing of winds from different evolved stellar evolutionary phases on the overall intracluster gas density and temperature.

Thus, the average mass lost by stars turning off the main sequence at time $t_i$ is given by a modified form of the methodology developed in \cite{Pooley2006}:
\begin{equation}
\langle \dot{M}_{\rm to} \rangle = \frac{ \langle \Delta M_{\rm to} \rangle }{t_i}
\label{eq:fulltomdot}
\end{equation}
and the average wind velocities are given by
\begin{equation}
\langle v_{\rm w}^2 \rangle = \frac{2\langle \Delta E_{\rm{K,to}}\rangle}{\langle \Delta M_{\rm to}\rangle}
\end{equation}
 where each average quantity is integrated from the zero age main sequence until the end of the RGB/AGB phases.
 Here, the IMF contribution from the turn-off stars folded into the expression defining $\langle \Delta M_{\rm to} \rangle$ \btt{(equation \ref{deltam2})}  gives the relative number of turn-off stars in a cluster. Thus, this mass loss rate is the normalized average {\bf{per star in the cluster}}.

To quantify the contribution to the mass-loss rates from stars with $M \approx M_{\rm to}$ in their post-main sequence evolutionary phase, we further quantize the RGB and AGB mass-loss phases as a function of the fraction of mass lost during a certain post-main sequence phase, $f_{j}$.  Here, we assume this fraction is counted from the end of the RGB/AGB phase back in time towards the turn-off.  In this way, $f_{j,\rm{min}}$ physically represents mass loss from a population of stars in which only one to tens of stars at the tip of the RGB/AGB contribute to the mass and energy injection into the cluster.  In contrast, $f_{j,\rm{max}}$ represents a population in which all evolved stars contribute to mass and energy injection, and the ejecta from the full population of evolved stars thermalizes and mixes efficiently.
For each time period $t_j$, we thus define \rt{the population fraction} $f_j$ as
\begin{equation}
f_j \equiv \frac{\langle \Delta M_{j,\star} \rangle}{\langle \Delta M_{\rm to} \rangle}
\label{eq:fjdef}
\end{equation}
such that the mass-loss rate for the population at time $t_j$ can be derived from equation \ref{eq:fulltomdot} as
\begin{equation}
\langle \dot{M}_j \rangle =  \frac{\langle \Delta M_{j} \rangle}{\Delta t_j}  = \frac{\langle \Delta M_{j,\star} \rangle}{\Delta t_j}  \left(\frac{N_j}{N_{\rm to}}\right)
 = \frac{f_j \langle \Delta M_{\rm to} \rangle}{\Delta t_j}   \left(\frac{N_j}{N_{\rm to}}\right)
 \label{eq:mdotfilong}
\end{equation}
where the time spent in each phase and the number of stars in each phase relative to the number of turn-off stars are given by $\Delta t_j = t_f - t_j$ and $\left( N_j / N_{\rm to} \right)$, respectively.  Assuming the number of stars in a phase $j$ is proportional to the time spent in this phase, $\left( N_j / N_{\rm to} \right) = \Delta t_j / \Delta t_{\rm to}$, reduces equation \ref{eq:mdotfilong} to
\begin{equation}
\langle \dot{M}_j \rangle = f_j \frac{\langle \Delta M_{\rm to} \rangle}{\Delta t_{\rm to}} = \frac{ f_j}{\Delta t_{\rm to}}  \left( \frac{N_{\rm to}}{N_{\rm tot}} \right) \int_{t_0}^{t_f}  \dot{M}(M_{\rm to},t) dt
\label{eq:mdotfi}
\end{equation}
where we have substituted $\langle \Delta M_{\rm to} \rangle$ from equation \ref{deltam2}, extending the integral over time from the turn-off, $t_0$ to the end of the RGB and AGB phases, $t_f$.

Using this expression for mass-loss at time $t_j$ and defining the wind speed from equation \ref{deltae} and
\begin{equation}
\left\langle v_{w,j}^2 \right\rangle = \frac{2 \langle \Delta E_{\rm{K,to}}(t_j) \rangle}{\langle \Delta M_j \rangle}
\end{equation}
the equation governing the wind speed of the stellar winds at time $t_j$ along the turn-off is
\begin{equation}
\left\langle v_{{\rm w},j}^2 \right\rangle = \frac{ \int_{t_j}^{t_f} \dot{M}(M_{\rm to},t) v_{\rm w}^2(M_{\rm to},t)dt}{f_j \langle \Delta M_{\rm to} \rangle}
\label{eq:vwfi}
\end{equation}

Individual mass-loss rates $\dot{M}(M_\star,t)$  and wind velocities $v_{\rm w}(M_\star,t)$ for each phase of stellar evolution are taken from the {\tt MIST-v1} stellar evolutionary models \citep{choi2016}.
The wind velocity is approximated to be equal to the escape velocity which is accurate to within a factor of a few across a wide range of masses and
life stages \citep{Abbott1978,Badalyan1992,Dupree1987,Loup1993,Vassiliadis1993,Schaerer1996,Nyman1992,Evans2004,Debes2006,Naiman2012}.
Stellar evolutionary models like the {\tt MIST-v1} models \citep{choi2016}, calculated with the {\tt MESA} stellar evolutionary code \citep{Paxton2011}, provide a reasonable estimate for both the mass-loss rates and wind velocities on both the main sequence and post-main sequence phases \citep{Naiman2012}.
While in practice we can define $f_{j,\rm{max}} = 1.0$ without issue, setting $f_{j,\rm{min}} = 0.0$ is equivalent to no mass being lost by any star as shown by equation \ref{eq:fjdef}, leading to an nonphysical form of equation \ref{eq:vwfi}.  \rt{From our particular choice of binning of $f_j$, i}n what follows, we assume $f_{j,\rm{min}}=2 \times 10^{-3}$ represents the mass loss from stars at the very tip of the RGB/AGB.  For a cluster of $10^5$ ($10^6$) stellar members, this is equivalent to assuming the mass and energy injection is dominated by the 3-5 (30-50) stars at the highest mass loss rates.

\begin{figure}
\centering \includegraphics[width=0.40\textwidth]{\figdir 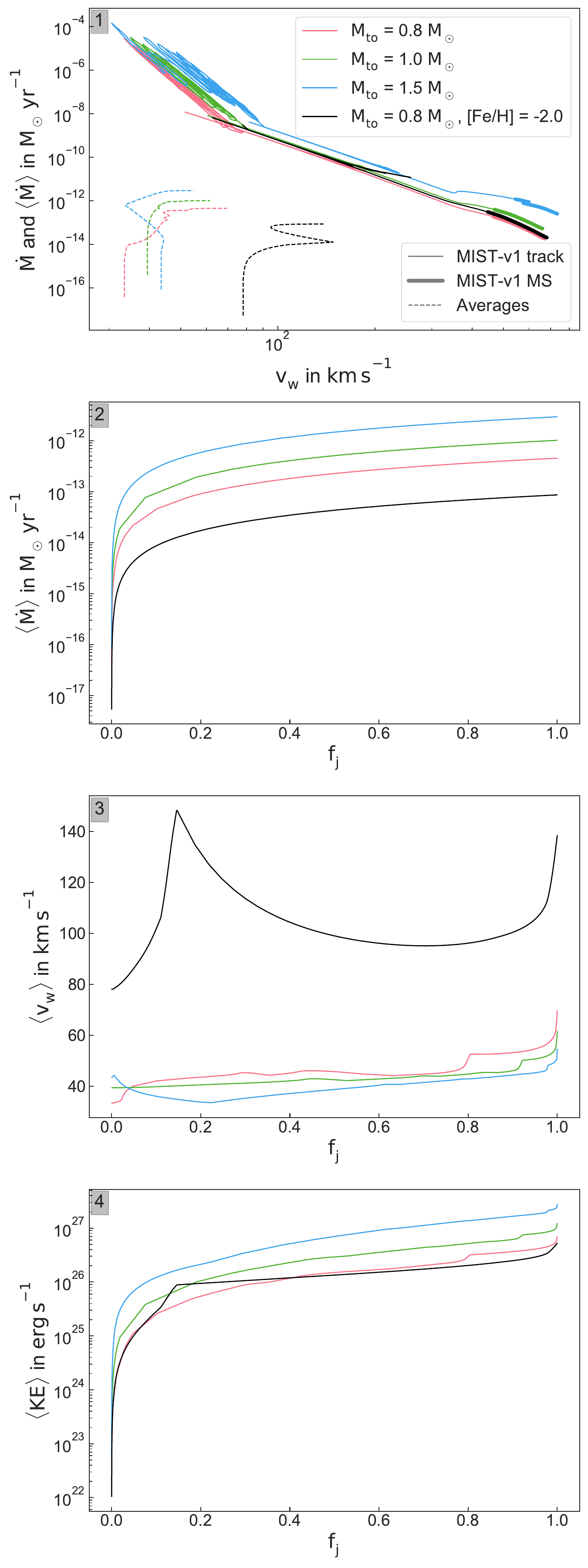}
\caption{Stellar mass-loss prescriptions for individual stars and our averaged quantities. {\it 1}: Solid colored lines show $\dot{M}$ -- $v_{\rm w}$ tracks for several {\tt MIST-v1} models \citep{choi2016} with metallicities of [Fe/H]$=-1.0$ (red, green, blue lines) and [Fe/H]$=-2.0$ (black line). The MS is highlighted with thicker lines.   Dashed lines show the averaged prescription from equations \ref{eq:mdotfi} and \ref{eq:vwfi} which \btt{are lower because of} the relatively few RGB/AGB stars losing mass at low $v_{\rm w}$ and high $\dot{M}$.  {\it 2}: The mass-loss rate as a function of the fractional mixing between different components of the evolved stellar population -- $f_j$ -- increases as more stars are added into the average.  While stars at the end of their lives, $f_j \sim 0$, have large mass-loss rates as shown in the upper left corner of the top panel, these stellar phases are short-lived and thus their average mass-loss rates are low.  {\it 3}: The average wind velocity for an evolved stellar population also increases as more stars are included in the averaging.  The lower metallicity model does not loose enough mass to pass through the high $\dot{M}$ and low $v_{\rm w}$ phase.   {\it 4}: The average kinetic energy injected increases with population fraction.  The low metallicity $0.8$~$\rm{M_\odot}$ model (black) injects a similar amount of energy as the high metallicity model (blue) at lower mass loss rates.}
\label{fig:newpre}
\end{figure}

The range of mass-loss and wind parameters from both the {\tt MIST-v1} models and our averaged prescription for several low-mass stellar models are shown in Figure~\ref{fig:newpre}.  Mass-loss and wind velocities are relatively slowly evolving quantities throughout the majority of a stars' lifetime on the main sequence. However, when stars traverse the RGB and AGB phases both their mass-loss rates and wind velocities vary dramatically over short timescales  for all but the lowest metallicity models as shown by the rapidly varying solid lines in the top panel of Figure~\ref{fig:newpre}. These large variations in the top panel of Figure~\ref{fig:newpre} at low $v_{\rm{w}}$ and high $\dot{\rm{M}}$ are the end of the star's life (RGB and AGB phases), unless the metallicity is low enough such that no thermally pulsing phase is present (dark solid line in Figure~\ref{fig:newpre}).  These short duration, large changes in wind velocity and mass loss rate are in significant contrast to the relatively stable wind velocities and mass loss rates present on the main sequence, as depicted by the thick solid lines at high $v_{\rm{w}}$ and low $\dot{\rm{M}}$ in the top panel of Figure~\ref{fig:newpre}.
The averaged mass-loss rates are significantly lower during these periods of rapid evolution as depicted by the dashed lines in the top panel of Figure~\ref{fig:newpre}.  This is due to the fact that the average mass-loss rates and wind velocities take into account the relatively few number of stars in these phases at any given time.  This is quantified further in the
two middle panels of Figure~\ref{fig:newpre} which show our parameterizations of mass-loss and wind velocity as a function of the mixing fraction between populations, $f_j$.  The mass-loss from a single star at the end of the RGB/AGB phase is given by $f_j \sim 0$, and while the rate of mass-loss from this single star is high, the short duration of this phase leads to a minimum averaged mass-loss rate as depicted in the second panel of Figure~\ref{fig:newpre}.  As $f_j$ increases and more stars at a variety of evolutionary phases are incorporated into the average, both the mass-loss rate and wind velocity increase as shown in the second and third panels of Figure~\ref{fig:newpre}, respectively.
Kinetic energy ejected by stellar winds per star increases as the population fraction increases as seen in the bottom panel of Figure~\ref{fig:newpre}.  The effects of metallicity on stellar wind ejecta are evident in Figure~\ref{fig:newpre} as well - the kinetic energy ejected by models with [Fe/H]=-1.0 and [Fe/H]=-2.0 are comparable though they have dissimilar mass loss rates and wind velocities at all population fractions.  We stress here that one could just as easily parameterize stellar wind mixing as fractions of total kinetic energy ejected as opposed to by total mass ejected as we have done in this paper, leading to slightly different curves than those presenting in Figure \ref{fig:newpre}.

\begin{figure}
\centering\includegraphics[width=0.48\textwidth]{\figdir 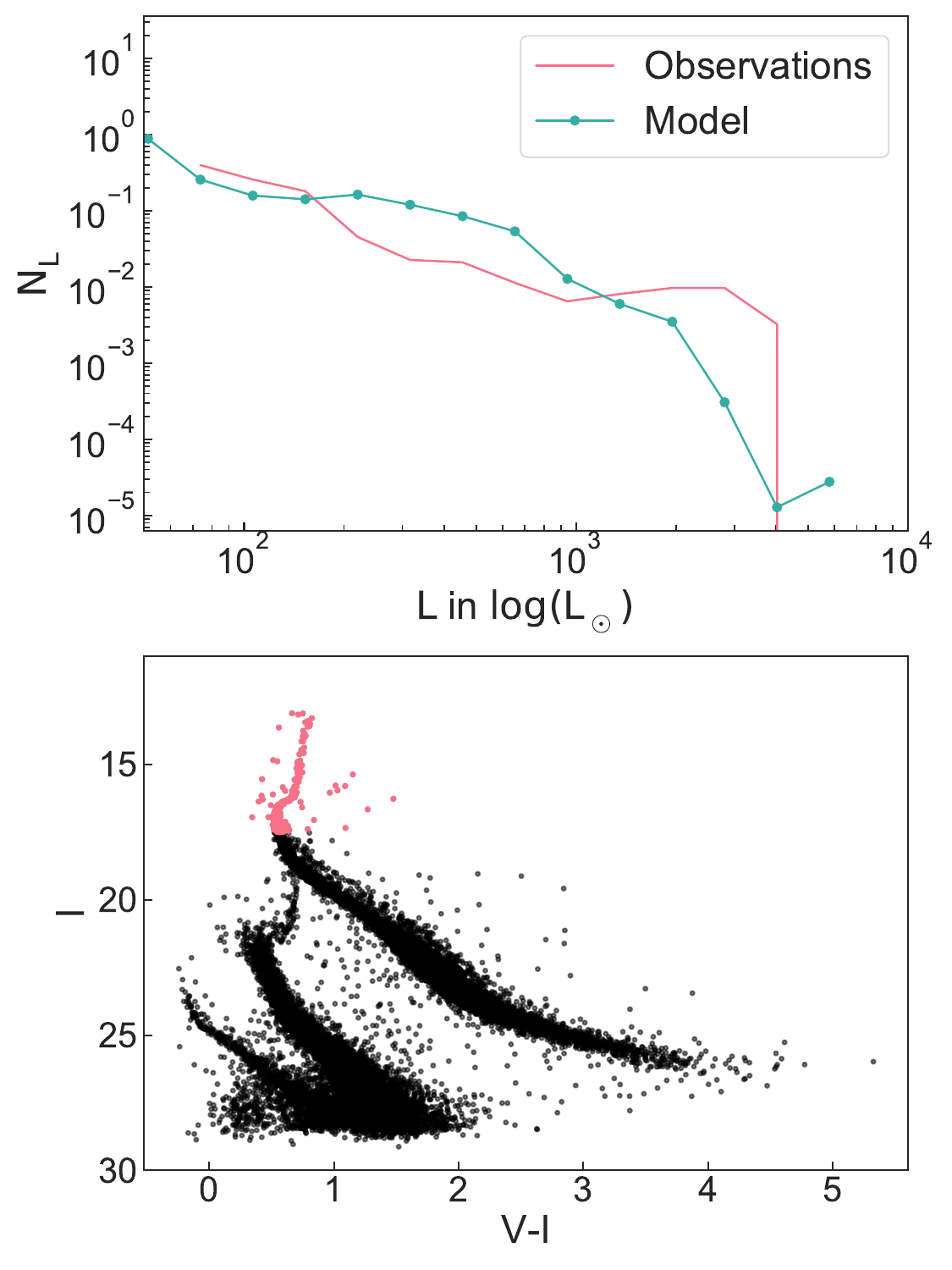}
\caption{The modeled number of stars with a given luminosity for a cluster with parameters similar to 47~Tuc.  Here, the {\tt MIST-v1} model for a \btt{star with} $\rm{M} = 0.85 \, \rm{M_\odot}$, [Fe/H]$ = -0.75$ is used -- similar to the turn-off mass and metallicity observed in 47~Tuc \citep{Harris1996a}.   The top panel shows the number of simulated turn off stars with a given luminosity (blue dots) and the histogram of turn-off and RGB/AGB stars with a given luminosity in 47~Tuc (pink line) from the observations of \citep{kalirai2012}.  The main sequence turn off stars selected from the observations of \citep{kalirai2012} are shown in the bottom figure as pink dots.  The bottom figure also shows the full main sequence as well as the white dwarf cooling sequence and stars in the background SMC with black dots.  Both datasets in the top panel have been normalized such that their number integrated over each luminosity bin equals the total number of stars in the pink highlighted region in the bottom plot -- approximately $620$ observed turn off stars.  Because stars at different phases (with different number counts) can have similar luminosities, the blue line \btt{in the top plot} represents the mean of the relation.}
\label{fig:ln}
\end{figure}

Given our prescription of the number of stars with $M\approx M_{\rm to}$ within the subpopulation described by $f_{ j}$,
 $\left( N_j/ N_{\rm{to}} \right) = \Delta t_j / \Delta t_{\rm to}$, for a given star cluster with a known turn-off mass we can estimate the number of stars with a given luminosity (the luminosity of stars in the bin $N_{\rm L} = N_{j+1}-N_{j}$) and compare this to observations.  Figure~\ref{fig:ln} shows these estimations for 47~Tuc with observed values of $M_{\rm{to}} = 0.86 \, \rm{M_\odot}$ \citep{thompson2010} and $\rm{[Fe/H]} = -0.72$ \citep{Harris1996a} using a {\tt MIST-v1} model with a similar mass and metallicity ($M_{\rm to} = 0.85$, $\rm{[Fe/H]} = -0.75$).   The luminosity function of simulated average stars roughly reproduces the observed distribution within the observational completeness limits.

\section{The Role of Stellar Wind Heating}

In what follows, we employ spherically symmetric one-dimensional hydrodynamical simulations to investigate the state of intracluster gas in present-day globular clusters.  Here, we assume the evolution of gas within these clusters is mediated by energy and mass  injection from stars at different post-main sequence phases, with the contribution of each phase quantified by the $f_j$ parameter.
The injection of energy by a much less abundant, yet more individually energetic, population of millisecond pulsars will not be examined until Section~\ref{sec: pulsarheating}.
Modeling the cluster masses, core radii, and velocity dispersions for specific clusters, we are able to compare our computational results to density constraints determined by observation.

\begin{figure}
\centering\includegraphics[width=0.42\textwidth]{\figdir 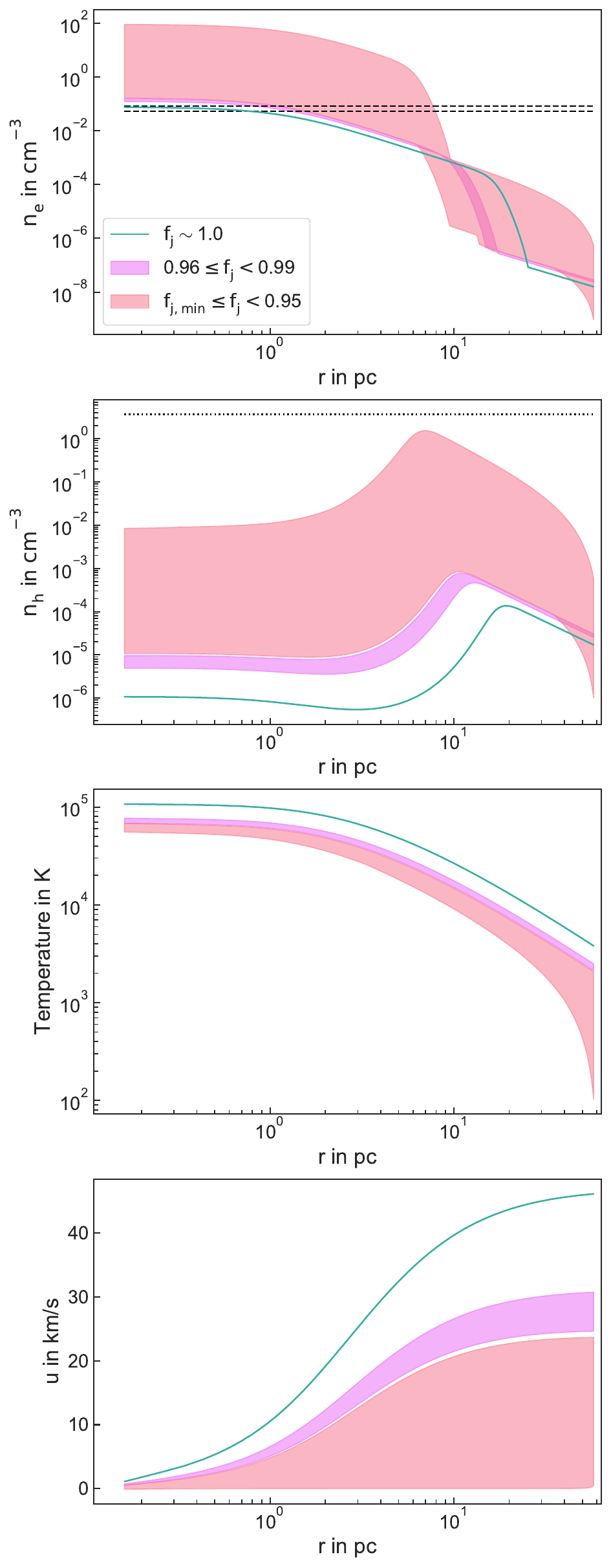}
\caption{ The state of the intracluster gas  in the globular cluster 47 Tucanae calculated using one-dimensional  hydrodynamical simulations.
The radial profiles of the electron  $n_{\rm e}$ and  neutral hydrogen $n_{\rm H}$  densities, the temperature $T$ and the flow velocity $u$ assumes energy and mass injection is determined solely by evolved stars.
The  stellar wind  thermalization and mixing efficiency  within the cluster between recent and late turn-off stars is regulated by $f_{ j}$, whose range is $\rm [0,1]$.
Electron constraints and neutral hydrogen upper limits are denoted by the \rt{horizontal} dashed lines and dotted line, respectively.
To model 47 Tucanae we employ a {\tt MIST-v1} stellar model with [Fe/H]$=-0.75$ and $M_{\rm to} = 0.85 \, {\rm M_\odot}$.  Only models with 0.96~$\le f_{ j} \le $~0.99  are are consistent with current observations, while models with less mixing produce either high ($f_{ j} \le $~0.95) or low ($f_{ j} = 1.0$) core-averaged electron densities (red, pink regions respectively).  The model sampling is $\Delta f_{ j} = 0.01$, thus all limits are approximate.}
\label{fig:47tuc}
\end{figure}

\begin{figure}
\centering\includegraphics[width=0.42\textwidth]{\figdir 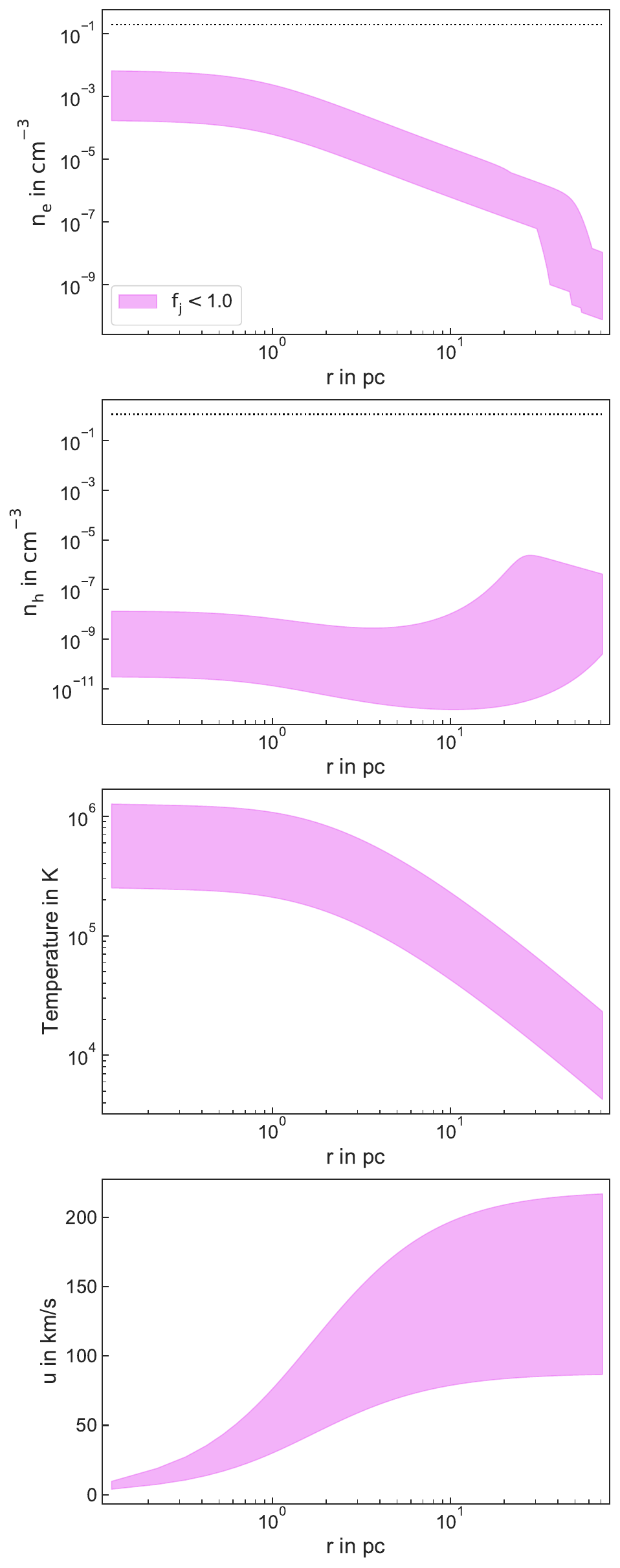}
\caption{ The state of the intracluster gas  in the globular cluster M15 calculated using one-dimensional  hydrodynamical simulations. Shown are the radial profiles  of the electron  ($n_{\rm e}$) and  neutral hydrogen ($n_{\rm H}$)  densities, the temperature $T$ and the flow velocity $u$.
The simulation assumes energy and mass  injection is determined solely by turn-off stars.
The  stellar wind  thermalization and mixing efficiency  within the cluster between recent and late turn-off stars is regulated by $f_{ j}$, whose range is $\rm [0,1]$.
The \rt{horizontal} dotted lines show the  electron and neutral hydrogen upper limits.
To model M15 we employ
a  {\tt MIST-v1} stellar model with [Fe/H]$=-2.50$ and $M_{\rm to} = 0.80 \, {\rm M_\odot}$.  All models have number densities below the observational limits, owing the fast winds of the lower metallicity stellar model and the smaller mass of M15 in comparison to 47~Tuc. }
\label{fig:m15}
\end{figure}

The simulation results for a cluster created to match the characteristics of 47~Tucanae are shown in Figure~\ref{fig:47tuc}.
We employed a cluster mass of $M_{\rm c} = 6.4 \times 10^5 \, {\rm M_\odot}$ \citep{Marks2010} and cluster velocity dispersion $\sigma_{\rm v} = 27 \, {\rm km\;{s^{-1}}}$ \citep{Bianchini2013}.
Stellar winds are taken from a model with a similar metallicity to M15, [Fe/H]$=-0.75$ \citep{Harris1996a}.
The resulting free electron density ($n_{\rm e}$), neutral hydrogen density ($n_{\rm H}$), temperature, and wind flow velocity profiles for the state of the intracluster gas in Galactic globular cluster 47~Tucanae are shown in Figure~\ref{fig:47tuc}, where dashed lines represent the free electron constraints and the neutral hydrogen upper limits taken from \citet{Freire2001b} and \citet{Smith1990}, respectively.
Here, electron and neutral hydrogen fractions are determined with the assumption that the gas is in collisional equilibrium, and thus, the electron gas fraction is dependent upon the temperature alone.  This amounts to solving for $n_{\rm e}$ and $n_{\rm H} = n_{\rm H_{\rm tot}} + n_{\rm H+}$, where $n_{\rm H+}$ is the ionized Hydrogen number density and $n_{\rm H_{\rm tot}}$ the total Hydrogen number density, in the collisional equilibrium equation: $\alpha_{\rm rec}(T) n_{\rm e} n_{\rm H+} = C_{\rm ci}(T) n_{\rm e} n_{\rm H_{\rm tot}}$, where the recombination coefficient, $\alpha_{\rm rec}(T)$, and the collisional ionization coefficient, $C_{\rm ci}(T)$ are functions of the temperature of the gas, $T$ \citep{Hummer1987,Padmanabhan2000}.
The three regions in this figure display models with mean electron densities higher or lower than the observed limits (\btt{large} red or \btt{small} blue regions) and models which fit the electron density observations (\btt{medium-sized} pink region).
Thus, in order to explain the low gas and dust levels observed in this cluster solely by the heating supplied from a variety of recent turn-off stars, winds from the entire population of stars, 0.96~$\le f_{ j} \le$~0.99 ($\left \langle v_{\rm w} \right \rangle \approx 54 \, \rm{km s^{-1}}$, $\left \langle \dot{M}\right \rangle \approx 5 \times 10^{-13} \, \rm{M_\odot yr^{-1}}$) must effectively thermalize and mix into the cluster environment.
Modeling 47~Tucanae is of significant importance, since after nearly half a century of searching,  the first ever detection of ionized intracluster gas took place here. This constraint strongly  limits the allowed values of  $f_{ j}$.

Similarly, the simulation results for cluster parameters set to match that of the Galactic globular clusters M15 are shown in Figure  \ref{fig:m15}.
A cluster mass of $M_{\rm c} = 4.4 \times 10^5 \, {\rm M_\odot}$ \citep{McNamara2004} and a cluster velocity dispersion of $ \sigma_{\rm v} = 28$ km s$^{-1}$ \citep{Harris1996a} was employed.
Stellar winds are taken from a model with a similar metallicity to M15, [Fe/H]$=-2.5$ \citep{Harris1996a}.
Figure  \ref{fig:m15} displays the free electron density, neutral hydrogen density, temperature, and wind flow velocity profiles, where dashed lines indicate the electron and neutral hydrogen upper limits from \citet{Freire2001b} and \citet{Anderson1993}, respectively.
All models result in electron and hydrogen densities that are consistent with observational constraints.  This is due to the combination of the lower metallicity and smaller mass of M15 in comparison to 47~Tuc -- lower metallicity results in faster stellar winds (Figure~\ref{fig:newpre}) and the smaller mass of M15 produces a gravitational potential which is slightly less conducive to retaining large central density reservoirs \citep{Naiman2012}.

In Figures \ref{fig:47tuc} and \ref{fig:m15}  we have examined the free electron density $n_{\rm e}$ profiles, however, we will be shifting our focus to  the average density within the cluster's core $\bar{n}_{\rm e}$ as we compare the observationally  allowed values of   $f_{ j}$  among different  globular clusters.  We have discussed the results of 47 Tucanae and M15, which happen to have similar velocity dispersions despite 47 Tucanae being  about 1.45 times more massive. Comparing the four clusters in our sample we see that  the velocity dispersion  has a larger spread in values, while the masses are more similar. Additionally, \citet{Naiman2011} found that for spherically symmetric simulated clusters, the cluster mass was less critical for gas retention when compared to the impact arising  from changes in the  velocity dispersion. Motivated by this, we have chosen to fix the cluster mass in an effort to systematically explore the effects of changing the velocity dispersion on the average free electron density $\bar{n}_{\rm e}$ and the average neutral hydrogen density $\bar{n}_{\rm H}$.

\begin{figure}
\centering\includegraphics[width=0.45\textwidth]{\figdir 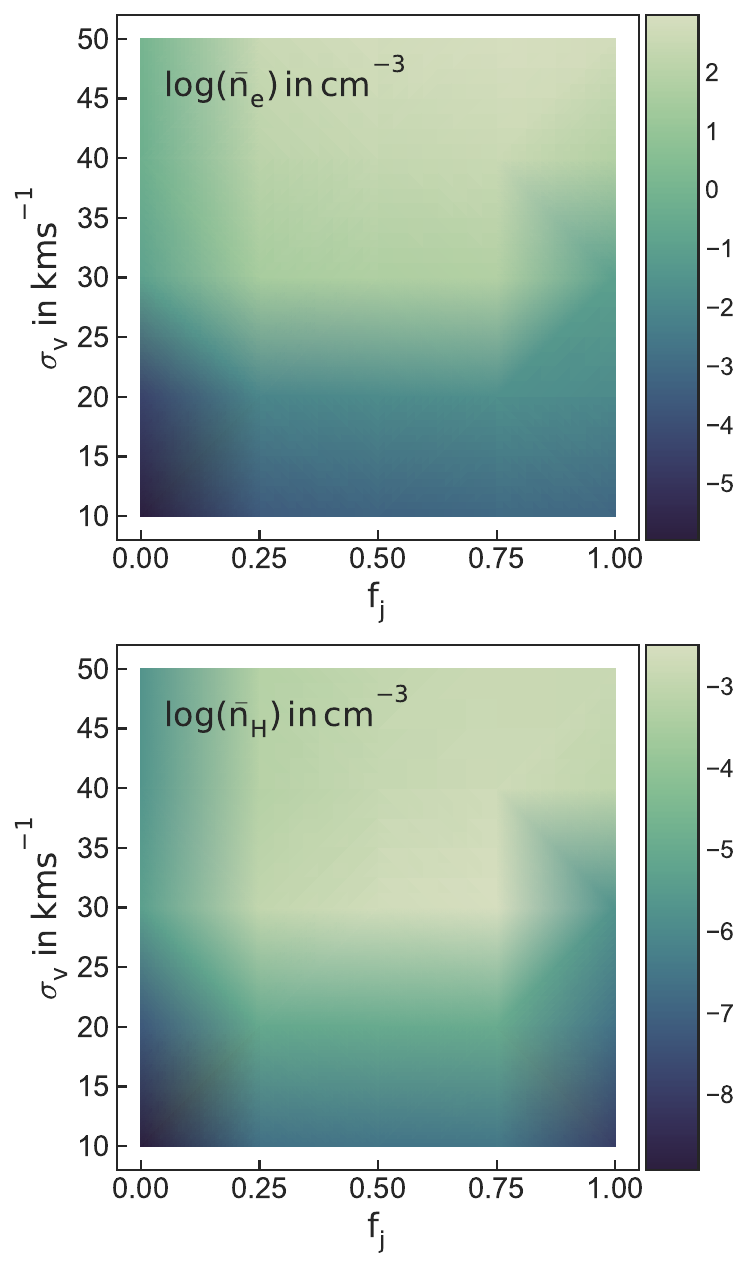}
\caption{ The relationship between a \btt{changing wind thermalization and mixing fraction $f_{\rm j}$ and cluster velocity dispersion $ \sigma_{\rm v}$ on} the free electron density $n_{\rm e}$  ({\it top panel}) and $n_{\rm H}$ ({\it bottom panel})  averaged over the core. In both panels, the cluster mass was held at a constant value, $M_{\rm c} =  5 \times 10^5 \, {\rm M_\odot}$. }
\label{fig:sigma}
\end{figure}

Figure~\ref{fig:sigma} displays the results of our effort to illustrate in more general terms how $\bar{n}_{\rm e}$ and $\bar{n}_{\rm H}$ change with velocity dispersion and $f_{ j}$ for a cluster of fixed mass\btt{, achieved in practice by varying the cluster radius}.
As expected, we see that, in general, a larger cluster velocity dispersion results in an increase in density for both $\bar{n}_{\rm e}$ and $\bar{n}_{\rm H}$.
At low velocity dispersions, the majority of the gas is easily removed from the cluster's potential regardless of the amount of recent turn-off star heating, leading to the low electron and hydrogen number densities for $\sigma_{\rm v} \lesssim 30 \, {\rm km\; s^{-1}}$.  As the velocity dispersion increases, more material is funneled toward the central regions of the cluster, and for high enough dispersions, this gas effectively cools as it collects in the cluster's core.
Enhancements in both electron and neutral hydrogen densities are maximized for $0.25 \lesssim f_{ j} \lesssim 0.75$.  For very low population fractions ($f_{ j} \lesssim 0.25$) there is not enough material ejected in stellar winds for large density enhancements to develop, while very large population fractions ($f_{ j} \gtrsim 0.75$) result in \rt{effectively} fast stellar winds which \rt{efficiently} eject gas from the clusters' centers.

\begin{figure}
\centering\includegraphics[width=0.45\textwidth]{\figdir 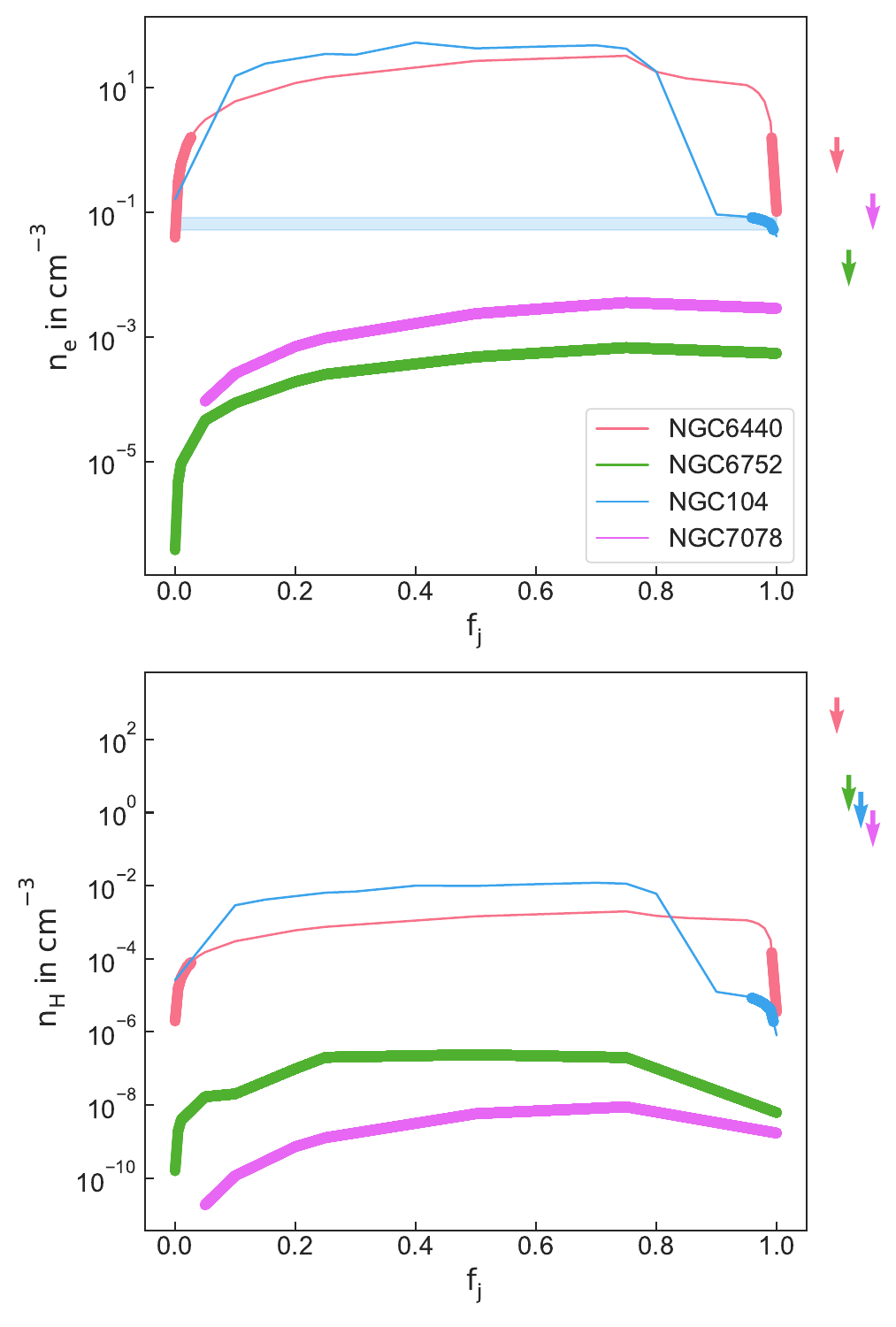}
\caption{ Free electron and neutral hydrogen densities, averaged over the cluster core, as a function of population mixing fraction for globular clusters 47 Tucanae, M15, NGC~6440, and NGC~6752. \btt{Observed upper limits on gas densities are shown with arrows.  The electron density limits of 47~Tucanae provide the tightest constraints and are shown by the shaded blue region in the upper panel \citep{Freire2001b}.} Values of $f_{ j}$ which produce densities consistent with the observed limits are highlighted with thick lines.
All of the profiles are consistent with the tightest population fraction of $f_{ j} \gtrsim $~0.95, provided by the 47 Tucanae electron density limits.
}
\label{fig:manyclusters}
\end{figure}

In Figure~\ref{fig:manyclusters}, we model the four clusters 47 Tucanae, M15, NGC~6440, and NGC~6752, where cluster masses and velocity dispersions are taken from \citet{Marks2010} and  \citet{Bianchini2013} for 47 Tucanae, \citet{McNamara2004} and \citet{Harris1996a} for M15, and \cite{Gnedin2002} for NGC~6440 and NGC~6752. With these parameters, we can explore the dependence of the free electron density and neutral hydrogen density, both averaged over the cluster core, on the population fraction.  The density constraints shown in Figure~\ref{fig:manyclusters} are  from \citet{Anderson1993}, \citet{Freire2001b}, \citet{Smith1990}, \citet{Hui2009} and  \citet{DAmico2002}.  The density profiles from each of the four clusters are roughly consistent with a population fraction of $f_{ j} \gtrsim $~0.95, determined by the detection of ionized gas in 47 Tucanae, and displayed as the shaded blue region in Figure~\ref{fig:manyclusters} \citep{Freire2001b}.

Both high and low $f_{ j}$ values are allowed for NGC6440 in Figure~\ref{fig:manyclusters}, with the constraint on the high end more stringent than that from NGC104 alone.
The strong turnover at low and high ends for the central $n_{\rm e}$ and $n_{\rm H}$ seen for NGC6440 is likely due to the high mass and velocity dispersion assumed for this cluster.   This compact potential aids in funneling large density enhancements into the cluster in all simulations except those with the least mass lost by stars, $f_{ j} \lesssim 0.02$ ($\left \langle v_{\rm w} \right \rangle \approx 35 \, \rm{km s^{-1}}$, $\left \langle \dot{M} \right \rangle \approx 8 \times 10^{-15} \, \rm{M_\odot yr^{-1}}$), and those with high wind velocities, $f_{ j} \gtrsim$~0.98 ($\left \langle v_{\rm w} \right \rangle \approx 54 \, \rm{km s^{-1}}$, $\left \langle \dot{M} \right \rangle \approx 5 \times 10^{-13} \, \rm{M_\odot yr^{-1}}$).
While these constraints seem stringent, they result from the sharp upturn in the wind velocity at large population fractions, as seen in Figure~\ref{fig:newpre}, and other prescriptions could lead to different constraints on $f_{ j}$.
The parameters governing the gravitational potentials and the limits on both free electron and neutral hydrogen densities used for all clusters simulated in this paper are collected in Table \ref{table:clusterparams}.

In summary, our models indicate that mixing between populations of evolved stars is essential in explaining the low density gas and dust observed in globular clusters and that a  minimum of about $\sim$95\% of the winds from the evolved stellar population must effectively be thermalized and mixed into the cluster environment to provide an accurate description of current observational constraints.
It is important to recognize that  $f_{ j}$  is  the fraction of the total amount of mass (and by extension in our prescription -- energy) injected by the different populations of evolved stars, which has been calculated here using  {\tt MESA}. Consequently, the  constraints we  have derived on $f_{j}$  are relative in the sense that their definition depends on the exact value of the total injected mass, which can depend on the specifics of the mass-loss prescription and stellar evolutionary models as a whole.

\begin{table*}
  \caption{Observed parameters for simulated clusters.}
  \label{table:clusterparams}
\scriptsize
    \begin{tabular}{lllllllll}
    \hline
Name & $M_{\rm c}$ [$\, {\rm M_\odot}$] & $\sigma_{\rm v}$ [$\rm{km s^{-1}}$] & [Fe/H] & $R_{\rm gc}$ [kpc] & $M_{\rm to}$ [$\, {\rm M_\odot}$] & Age [Gyrs] & $e^-$ limit & H limit \\ \hline\hline
NGC104 (47~Tuc)  & $6.4 \times 10^5 \, {\rm M_\odot}$\textsuperscript{[1]} & 27\textsuperscript{[2]} & -0.72\textsuperscript{[3]} & 7.4\textsuperscript{[3]} &  0.86\textsuperscript{[4]} & 11.6\textsuperscript{[5]} & $\rm{n_e = 0.067 \pm 0.015 \, cm^{-3} \,  }$\textsuperscript{[6]}  & $\rm{M_H \le 3.7 \, {\rm M_\odot}}$\textsuperscript{[7]}  \\
NGC7078 (M15)  & $4.4 \times 10^5 \, {\rm M_\odot}$\textsuperscript{[8]} & 28\textsuperscript{[8]} & -2.37\textsuperscript{[3]} & 10.4\textsuperscript{[3]} &  0.80\textsuperscript{[9]} & 13.7\textsuperscript{[10]} & $\rm{n_e \le 0.2 \, cm^{-3} \,  }$\textsuperscript{[11]}  & $\rm{M_H \le 0.3 \, {\rm M_\odot}}$\textsuperscript{[12]}  \\
NGC6440  & $8.1 \times 10^5 \, {\rm M_\odot}$\textsuperscript{[13]} & 32\textsuperscript{[13]} & -0.36\textsuperscript{[3]} & 1.3\textsuperscript{[3]} &  0.85\textsuperscript{[*]} & 11.0\textsuperscript{[14]} & $\rm{n_e \le 1.6 \, cm^{-3} \,  }$\textsuperscript{[11]}  & $\rm{n_{\rm H} \le 5.9 \times 10^{21} \, cm^{-1}}$\textsuperscript{[15]}  \\
NGC6752  & $3.1 \times 10^5 \, {\rm M_\odot}$\textsuperscript{[13]} & 32\textsuperscript{[13]} & -1.54\textsuperscript{[3]} & 5.2\textsuperscript{[3]} &  0.80\textsuperscript{[*]} & 12.1\textsuperscript{[5]} & $\rm{n_e \le 0.025 \, cm^{-3} \,  }$\textsuperscript{[16]}  & $\rm{n_{\rm H} \le 2.2 \times 10^{20} \, cm^{-1}}$\textsuperscript{[15]}  \\
\hline
    \end{tabular} \\
   {\begin{flushleft}References: [1] \cite{Marks2010}, [2] \citet{Bianchini2013}, [3] \citet{Harris1996a}, [4] \citet{thompson2010}, [5] \citet{corr2016}, [6] \citet{Freire2001b},  [7] \citet{Smith1990},
   [8] \citet{McNamara2004}, [9] \citet{fahlman1985}, [10] \citet{monelli2015}, [11] \citet{Freire2001b}, [12]  \citet{Anderson1993}, [13] \citet{Gnedin2002}, [14] \citet{origlia2008}, [15] \citet{Hui2009}, [16] \citet{DAmico2002},
   [*] Estimated from overall GC averages.
   \end{flushleft}}
\end{table*}

\section{The role of pulsar Heating: The case of 47 Tucanae}
\label{sec: pulsarheating}

The role of pulsar heating in the evacuation of gas and dust in globular clusters was first discussed by  \citet{Spergel1991}.
At that time the total number of detected millisecond pulsars residing in globular clusters was about two dozen. We now know of over 150 millisecond pulsars in 28 separate globular clusters \citep{Freire2013}. In some of these clusters the population of millisecond pulsars is considerable. Globular cluster Terzan 5 is known to harbor 38 millisecond pulsars and 47~Tucanae contains 25 detected pulsars\footnote{\url{http://www.naic.edu/~pfreire/GCpsr.html}}.
In our hydrodynamical models, 47~Tucanae is used as a proxy to explore the physics of gas retention in clusters hosting a population of millisecond pulsars. This cluster was chosen because of  the  detection of ionized intracluster material \citep{Freire2001b}, which  allows for  strict  constraints on the efficiency of pulsar heating to be placed.

The hydrodynamical influence of millisecond pulsars in our simulations is restricted  to energy injection, $q_{\varepsilon,\Omega}(r,t)$, as the stellar winds  dominate the mass supply. For simplicity, we assume that the pulsar energy injection follows the spin down luminosity distribution -- the loss rate of rotational energy which is assumed to be equal to the pulsars' magnetic dipole radiation --  which here we parameterize as $q_{\varepsilon, \Omega}(r) = L_{\rm 0}/(r+r_{\rm L})^\beta$ where $L_{\rm 0}$, $r_{\rm L}$ and $\beta$ are derived from the observed spin down luminosity profile of 47~Tuc.  In fitting to the observed distribution, we impose
\begin{equation}
4 \pi \int_{r_{\rm min}}^{r_{\rm max}} q_{\varepsilon, \Omega}(r) r^2 dr = \sum_i L_{{\rm p},i}
\end{equation}
where $r_{\rm min}$ and $r_{\rm max}$ are the minimum and maximum \btt{cluster-centric} radii for pulsars observed in a cluster and $L_{{\rm p},i}$ are the individual spin down luminosities for each pulsar in the cluster.  We calculate $L_{{\rm p},i} = \left | -4 \pi^2 10^{45} \dot{P}/P^3 \right | \, \rm{ergs \, s^{-1}}$ with period $P$ and only including pulsars with spin down rates, $\dot{P} < 0$ \citep[e.g.][]{ostriker1969,finzi1969,shapiro1983}.

 As pulsar energy is predominately supplied  as Poynting flux,  the thermalization and mixing efficiency within the cluster  core remains highly uncertain.  In addition, also uncertain is the total amount of energy, $\dot{E}_{\Omega} = 4 \pi \int_{r_{\rm min}}^{r_{\rm max}} q_{\varepsilon, \Omega}(r) r^2 dr$, injected into the cluster. This uncertainty is largely due to unreliable timing solutions. For this reason,
the total energy injected by the pulsars that  is effectively thermalized and mixed into the cluster gas, $\dot{E}_{\rm p}$,  is treated as a free parameter and is  parameterized here \btt{with the pulsar energy thermalization and mixing fraction, $f_{\rm p}$, such that} $\dot{E}_{\rm p}=f_{\rm p} \dot{E}_{\rm \Omega}$.

As of the date that our simulations were run, there are 13 millisecond pulsars within 47 Tucanae with known spin-down timing solutions. If we, for example,  take the timing solutions at face value \citep{Manchester1990, Manchester1991, Robinson1995, Camilo2000, Edmonds2001, Freire2001c, Edmonds2002, Freire2003, Lorimer2003, Bogdanov2005}, ignoring the likely possibility that they might be corrupted by cluster motions, we find a total spin-down luminosity of $\dot{E}_{\Omega}=8.4 \times 10^{35}\;{\rm erg\;s^{-1}}$.  For comparison, using the mass-loss values depicted in Figure~\ref{fig:newpre}, $\dot{E}_{\Omega}$ is roughly four orders of magnitude larger than the stellar wind luminosity ejected by a $f_{ j} = 1.0$ population of evolved stars.  If, on the other hand, we assume that the luminosity distribution of millisecond pulsars residing in the cluster is well described by the luminosity function of  Galactic field pulsars \citep{Manchester2005},  we obtain $\dot{E}_{\Omega} = 5 \times 10^{34}\;{\rm erg\;s^{-1}}$, which is an order of magnitude smaller than the total power estimated using 47 Tucanae's pulsar timing parameters.

\begin{figure}
\centering\includegraphics[width=0.44\textwidth]{\figdir 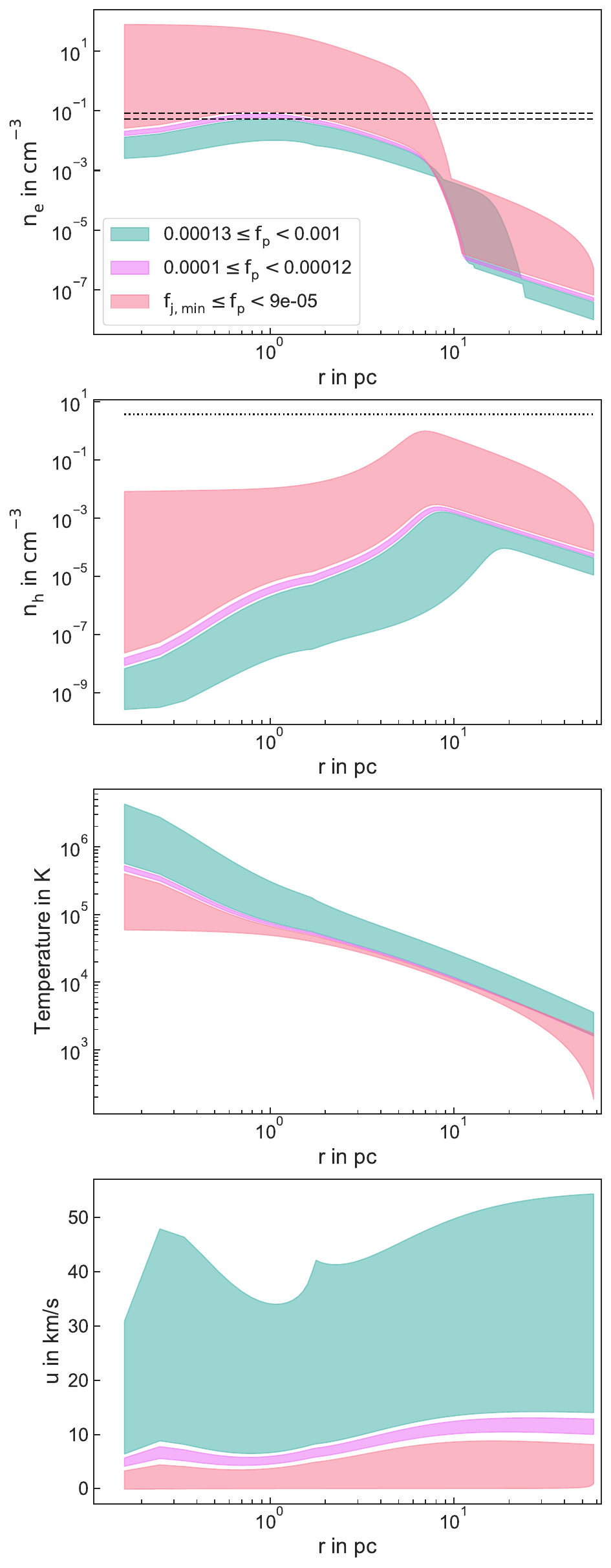}
\caption{ The state of the intracluster gas in the globular cluster 47 Tucanae as influenced by stellar winds from evolved stars with a population mixing fraction of $f_{ j} = 0.75$  and heating from the millisecond pulsar population. Shown are the radial profiles of the free electron density $ n_{\rm e}$, neutral hydrogen density $ n_{\rm H}$, temperature $T$, and flow velocity $u$, calculated using one-dimensional hydrodynamical simulations.  Models consistent with observations have pulsar heating fractions of 0.0001$\le f_{\rm p} \le$0.00012 and are shown in pink, while others produce central density enhancements which are either too high (red region, $ f_{\rm p} \le$9e-5) or too low (blue region, $ f_{\rm p} \ge$0.00013). \rt{Electron constraints and neutral hydrogen upper limits are denoted by the horizontal dashed lines and dotted line, respectively.}
}
\label{fig:compareHeating}
\end{figure}

Figure~\ref{fig:compareHeating} shows the free electron density, neutral hydrogen density, temperature, and flow velocity radial profiles for a cluster modeled after 47 Tucanae when heating from the millisecond pulsar population is included.  With the wind population fraction is fixed at $f_{ j} = 0.75$ as it is in Figure~\ref{fig:compareHeating}, models with 0.0001$ \le f_{ p} \le$0.00012 are consistent with free electron and neutral hydrogen density constraints.
Other models shown in Figure~\ref{fig:compareHeating} result in gas that is either too cold and dense  (red region) or too hot and diffuse (blue region) when \rt{compared}  to  observations.
As for the $f_{ j} = 0.75$ model ($\langle \dot{M} \rangle \approx 3.8 \times 10^{-13} \, {\rm M_\odot} yr^{-1}$, $\langle v_{\rm w} \rangle \approx 44 \, {\rm km s^{-1}}$), the rate of kinetic energy injected from winds alone is approximately a factor of $5 \times 10^{-4}$ lower than that ejected by the pulsars ($\dot{E}_{\Omega}$)
we conclude that the \btt{efficiency of the} pulsar energy injection \btt{in heating the intracluster gas} needs to be lower than the stellar wind contribution and, as a result, the currently poorly understood pulsar wind thermalization efficiency within the cluster's core must be small. This can be clearly seen in Figure~\ref{fig:compareHeating} by  comparing the thermodynamical profiles generated by models that include high levels of pulsar heating with those that include small levels. For reference, on the condition that the millisecond  pulsars inject  $5 \times 10^{34} {\rm erg\;s^{-1}}$, as inferred from the Galactic  field population, the thermalization efficiency  needs to be $\lesssim$0.002. \btt{This is in stark contrast to measurements approximating the thermalization fraction of the Crab nebula to be $\approx$10\% \citep[e.g.][]{smith2013} and models of pulsar-powered supernovae in which it is generally assumed that the total magnetar energy is thermalized throughout the remnant \citep{kasen2010}.}

Finally, it is worth noting here that as total mass-loss rates and wind velocities ejected by the stellar population are sensitive to the mass and metallicity of the {\tt MIST-v1}  model used so too will be the exact fraction of pulsar heating needed to evacuate the intracluster gas.

\begin{figure}
\centering\includegraphics[width=0.45\textwidth]{\figdir 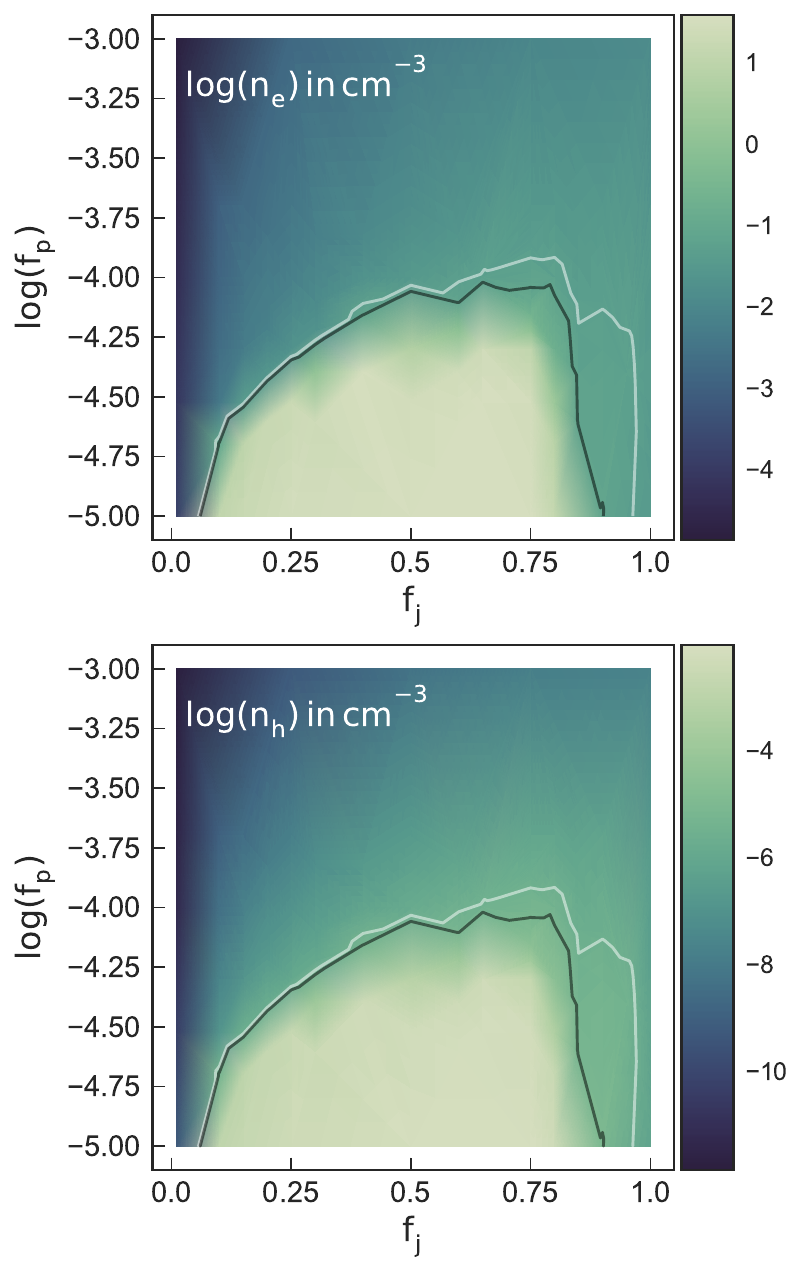}
\caption{The state of the free electron density $ n_{\rm e}$ and neutral hydrogen density $n_{\rm H}$ averaged over the core of the cluster as a function of millisecond pulsar heating and energy injection from the stellar winds for globular cluster 47 Tucanae.  The models between the white and gray lines in the $n_{\rm e}$ plot denote regions where the simulated $n_{\rm e}$ and $n_{\rm H}$ are within the density constraints. The region enclosed by the gray line are models which have $n_{\rm e}$ greater than the limits, while models outside the white line have $n_{\rm e}$ less than the limits.  All models produce $n_{\rm H}$ underneath the observed limit.
As our models do not cover the entire $f_{ j}$ and $f_{ p}$ parameter space, these regions are interpolated from the modeled parameter space. Note that our model places stringent  constraints on the thermalization efficiency of pulsar winds in the core of  47 Tucanae.  }
\label{fig:47tucConstraints}
\end{figure}

In Figure~\ref{fig:47tucConstraints} we consider  a more comprehensive scenario in which  heating from both stellar winds and the  millisecond pulsar populations are varied. In this case, we search for models that  produce thermodynamical profiles with average central electron and neutral hydrogen densities that are consistent with observational constraints  when energy injection  from both pulsars and  stellar winds is  taken into account.
The \btt{region} between the \btt{black and white lines} show\btt{n} in the upper $n_{\rm e}$ plot of Figure~\ref{fig:47tucConstraints} denotes the parameter space that \btt{falls} within the density constraints for 47 Tucanae. It is important to note the low levels of millisecond pulsar energy thermalization  needed to explain the  density limits  even when the \btt{wind} population fraction, $f_{ j}$, is low \btt{-- at most a thermalization of $\sim 0.01\%$ of the total cluster pulsar spin-down luminosity is required to explain the observed low levels of gas in 47~Tucanae}. This is because of the overabundance of millisecond pulsars residing in the core of  47 Tucanae, which helps prevent the intracluster gas from being effectively retained in the cluster \btt{by concentrating the deposition of energy were the densest gas resides}.  As such, we \btt{are not only able to derive that pulsar spin-down luminosity can efficiently eject material from a cluster if a small fraction of this energy is thermalized effectively in the gas, we can further} conclude that current observations place strong constraints on  the ability of pulsar winds to effectively  thermalize and mix  within cluster's cores.

\section{Discussion}

In this paper, we examine several gas evacuation mechanisms in an effort to account for the paucity of gas and dust in globular clusters.
The tenuity of the intracluster medium is observed consistently from cluster to cluster and, as such, we aim to distinguish a mechanism that is universal in scope and not specific to variable cluster properties (e.g. UV heating from the HB stars \citep{Vandenberg1977},  stellar collisions \citep{Umbreit2008}).
Energy injection by various subpopulations of evolved stars within a cluster has generally been dismissed, primarily due to the fact that the energy contribution per star is low when compared to explosive processes, such \rt{as} hydrogen rich novae \citep{Scott1978,Moore2011}.
Here, we argue that the sheer abundance of the less evolved turn-off stellar members warrants this mechanism worthy of consideration.
To this end, we construct  one dimensional hydrodynamical models to study the properties of the gas in  globular clusters with mass and energy  injection provided by stars in different phases of their post-main sequence evolution.
Choosing  our initial conditions  to match  the cluster masses and core radii of globular clusters  47 Tucanae, M15, NGC~6440, and NGC~6752, we are able to compare our simulation results with observational density constraints.
Using our wind thermalization prescription we find that a minimum of approximately $f_{ j} \gtrsim$ 95\% (see section \ref{sec:se}) of the total stellar wind luminosity from the entire population of evolved stars,  which we  calculate using  {\tt MIST-v1} stellar evolution models, must be effectively thermalized and mixed into the cluster medium in order to generate results that are in agreement with current observational limits.
We conclude that the energy output from the evolved stellar population alone is capable of effectively  sweeping out the highly evolved massive stellar ejecta in all the systems we have modeled. Specifically, we argue this result distinguishes a viable ubiquitous gas and dust evacuation mechanism for globular clusters.
It is important to note that  given current  mass-loss uncertainties, and their dependence on the exact mass and metallicity of the star modeled, it is difficult to precisely quantify the exact population fraction.  However, it is clear that  on the basis of commonly used  mass-loss  rate prescriptions, we expect energy injection from post-main sequence stars  to play a  vital role in regulating gas retention in globular clusters.

We extend our computational analysis to investigate the efficiency of pulsar wind feedback in a simulation modeled after the globular cluster 47 Tucanae, which is known to harbor 25 millisecond pulsars. The detection of ionized intracluster gas within 47 Tucanae allows for a detailed comparison between simulated results and the strict observational density constraints.
The millisecond pulsar energy injection is known to be rather significant  and, as such, we conclude that the pulsar wind thermalization efficiency must be extremely low in order to maintain  the low density constraints for this  cluster.
Other clusters of interest in our analysis, M15, NGC~6440, and NGC~6752, are known to host smaller populations of 8, 6, and 5 millisecond pulsars, respectively. While there is a high variability in the total pulsar energy injection per cluster, all observations \btt{of gas plasma densities made with these pulsar populations} indicate a tenuous intracluster medium.

\btt{It is likely that, when present, the millisecond pulsar population thermalizes gas inefficiently, requiring only a small fraction of the spin-down luminosity to be mixed within the gas to drive central densities to their low observed values.}
The heat supplied by millisecond pulsars within a globular cluster is difficult to estimate,  mainly due to the highly uncertain  thermalization and mixing efficiency  of the emanating  Poynting flux  within the core of the cluster.
What remains to be demonstrated, perhaps by means of three-dimensional, magneto-hydrodynamical simulations, is that  pulsar outflows are not efficiently  polluted by  the baryons emanating from the evolved stars by the time that they reach the edge of the cluster. We suspect, based on current observational constraints,  that the pulsar wind energy is only efficiently thermalized at much larger radii.
Observations of globular clusters with accompanying X-ray haloes support this idea \citep{Mirabal2010}. In addition, it has been suggested by \citet{Hui2009} that while the luminous X-ray pulsar wind nebulae have been detected from pulsars in the Galaxy, there is no evidence of a contribution to the diffuse X-ray emission by pulsar wind nebulae within globular clusters.  Searching for  low and high energy  diffuse emission within and around  pulsar-hosting globular clusters could, in principle, help \rt{uncover} the heating  structures from these objects and provide a much clearer understanding of the underlying processes at work.

The  modeling of  mass retention in globular clusters  continues to be  a formidable challenge to observers and theorists. The best prospects  lie with performing three-dimensional  magnetohydrodynamical simulations, probing the interaction of main sequence winds, evolved stellar winds, and,  when present in sizable numbers, pulsar winds.
While it is observationally challenging to detect signatures of the intracluster medium in extremely diffuse environments, forthcoming space and ground-based observations should provide the evidence necessary to unveil the detailed nature of the intracluster gas.

\section*{Acknowledgements}

Support  was provided by the Packard Foundation,  UCMEXUS (CN-12-578),  NSF AARF award AST-1402480, a NSF GRFP, and the Institute of Theory and Computation at the Harvard-Smithsonian Center for Astrophysics.
We thank the Aspen Center for Physics for its hospitality during the completion of this work.

\bsp	





\bibliographystyle{mnras}

\bibliography{gcp_paper}


\label{lastpage}
\end{document}